\documentstyle[osa,manuscript]{revtex}
\input{psfig}

\begin{document} 
\begin{titlepage}	
\begin{center}
{\Large {\bf 
Softening of Phase Transition in 2D Potts \\ 
Model Under Quenched Bond Randomness }}
\end{center}
\vskip 1.5cm
\centerline{ Fatih Ya\c{s}ar, Yi\u{g}it G\"{u}nd\"{u}\c{c} 
and Tar{\i}k \c{C}elik }
\centerline{\it Hacettepe University, Physics Department, Beytepe 06532, Ankara }
\vskip 2.5cm

{\small 
We have simulated, by using cluster algorithm, the $q=8$ state 
Potts model in two-dimension with varying amount of quenched bond randomness.
We have shown that there exist a finite size dependent threshold value
of the introduced quenched bond randomness for rounding  the first-order
phase transition and this threshold  value becomes smaller as the system size
increased.
\vskip 0.5cm
}



\end{titlepage}

The effects of the addition of quenched bond randomness to classical systems
whose pure version displays a continuous phase transition is well understood
in terms of the Harris conjecture \cite{Harris:1974}, which states that the values of the 
critical exponents change only if the specific heat exponent $\alpha$ is 
positive. Recently the introduction of randomness to the systems undergoing 
a first-order phase transition has attracted much interest.  Phenomenological 
renormalization group  arguments by Hui and Berker  \cite{Hui:1989} and rigorous proof of 
the vanishing of the latent heat by Aizenman and Wehr \cite{Aizenman:1989} showed that bond
randomness will induce a second-order phase transition in a system that would
otherwise undergo a symmetry-breaking first-order phase transition. The 
renormalization group (non-rigorous) arguments state that any infinitesimal amount of 
bond randomness will drive the system to possess a second-order phase 
transition for $d \le 2$ where d is the dimension of space. The numerical
evidence for this prediction is provided by Monte Carlo simulation of
two-dimensional 8-state Potts model \cite{Potts:1952} under a certain type of bond
randomness \cite{Landau:1992}. While the pure model is known to possess a strong first-order
phase transition, the simulations with two sets of randomly distributed
ferromagnetic bonds of two different strengths of certain ratio gave
clear evidences of a second-order transition with 2D Ising exponents.
For the convenience of establishing the histograms, the strong to weak bond 
ratio was chosen as $r=2$ or 10 in Ref. 5 and the simulations were carried out on
lattices of linear size $12 \le L \le 128$. In this context, it would be
quite relevant to study the changes in the characteristic behaviors of the 
system with respect to the introduction of a gradually increasing degree of 
bond randomness  as well as its finite size dependence.

Distinguishing the order of a phase transition is one of the major
problems concerning computer simulation studies of spin systems.  The major
difficulty, in this respect, arises when the correlation length is
finite but larger than the lattice size. In such situations, commonly
used tools of identifying the order of a transition (e.g. by looking
for the minima in free energy \cite{Lee:1991} or by
considering the probability distributions of energy
\cite{Challa:1986}) may not be indicative.  Even if there exist
metastable states, the size of the system may prevent the observation
of the double-peak behavior in the energy distribution.  The basic
reason for such  behavior may be that energy is a local
quantity. Quantities of global nature are expected to be more
sensitive to the correlation length and hence the effects of having
metastable states may be more emphasized.
In an earlier work \cite{Aydin:1996,Gunduc:1996} on two-dimensional
Potts model, it was observed that compared to the local operators such
as energy and order parameter, cluster related (global) operators are
more sensitive to structural changes in the system going through a phase
transition. Particularly, the average cluster size distribution may
give better indication on the order of the phase transition at smaller
lattice sizes than that for the energy distribution.  
The average cluster size can be defined as
\begin{equation}
ACS = {1 \over N_C} < \sum_{i=1}^{N_C} C_i>
\end{equation}
where $N_C$ is the number of clusters and $C_i$ is the number of spins in 
the $i^{th}$ cluster normalized by dividing by the
total number of spins.

$\;$ 

The aim in this work is to simulate, by using cluster algorithm, the 
two-dimensional $q$-state Potts model with a varying quenched bond randomness
and to observe the energy and the average cluster size histograms in
order to gain insight concerning the order of the phase transition in
the system.

$\;$

The Hamiltonian of the Potts model~\cite{Wu:1982} is given by 
\begin{equation}
     {\cal H} = \sum_{<i,j>}  K_{ij} \delta_{\sigma_{i},\sigma_{j}}.
\end{equation}
Here $K_{ij}=J_{ij}/kT$ ; where $k$ and $T$ are the Boltzmann constant and the
temperature respectively, and $J_{ij}$ is the magnetic interaction between spins
$\sigma_{i}$ and $\sigma_{j}$, which can take values $1,2,\dots, q$ for the
$q$-state Potts model. In the pure system, $K_{ij}$ is a constant and the 
value of the transition temperature is known for all q as 
$K_c = ln( 1 + \sqrt {q})$. In the random bond Potts model, the couplings
are selected from two positive values with a weak to strong ratio
$r = K_w / K_s$  which are chosen randomly from the distribution
$$P(K) = p \delta(K - K_w) + (1 - p) \delta(K - K_s).$$
When the probability p is taken as $p = 0.5$, the duality relation 
~\cite{Kinzel:1981}
$$(e^{K_w ^c} - 1)(e^{K_s ^c} - 1) = q$$
gives the critical values $K_w ^c$ and $K_s ^c$ of the corresponding
couplings  $K_w$ and  $K_s$.

$\;$ 

The two-dimensional $q$-state Potts model is known to undergo a
first-order phase transition for $q \ge 5$ ~\cite{Baxter:1973}.  
In our simulation of the $8-$state Potts model, the measurements are done
on three lattices of  the sizes  $32^2, 64^2$ and $128^2$. Since the
calculated correlation length $\xi$ of this model is about 20 lattice
sites \cite{Buffenoir:1993}, these lattices are considered to be
an indicative set for both large lattice measurements and for the
lattices of the order of the correlation length. Measurements are done
about the transition temperature and at each temperature $5 \times
10^5$ iterations are performed following the thermalization runs of $5
\times 10^4$ to $10^5$ iterations. For all simulation works, the cluster
update algorithm ~\cite{Swendsen:1987,Wolff:1989} is employed.

$\;$ 

We have examined the phase structure of the 8-state Potts model with
respect to the quenched randomness in the range starting from  $r = 1$ 
(the pure case) to $r = 0.1$ (weak bonds are ten times weaker). 
For a fixed value of r, twenty replicas are created by randomly distributing 
the strong and weak bonds over the lattice. The histograms for energy and 
the average cluster size are obtained by averaging the corresponding
quantities over the replicas. In Fig.1, we show the energy histogram on
$64 \times 64$ lattice for several values of r. Starting from the pure
case which exhibits a rather strong first-order phase transition, the 
8-state Potts model with quenched bond randomness displays a 
double-peak structure in the energy histogram down to $r = 0.5$, the latter
value corresponds to the case studied in Ref. 5.
By further tuning, one observes a single-gaussian in the energy 
histogram for $r = 0.46$,
which Gaussian gets sharper as the value of r chosen smaller.
In order to obtain a better insight concerning the order of phase
transition, we examined the average cluster size distributions which was
found to yield quite useful information concerning the order of the phase
transition in pure q-state Potts model phase \cite{Aydin:1996,Gunduc:1996}.
As shown in Fig. 2, the apparent peaks in the small cluster region,
 which is an indication of 
having metastable state disappear for $r = 0.46$. The phase transition
studied here on $64 \times 64$ lattice changes from first-order to
second-order somewhere in the range $0.46 \le r \le 0.50$, the exact
value of which is not attempted here to be evaluated.  The observed threshold 
value for the randomness is   specific to the lattice size used.
Therefore the next step is to investigate how this threshold value changes
with the size of the lattice. Fig. 3 displays the energy histograms for a
fixed value of r, namely $r = 0.53$, obtained on different size lattices. Here, this
value of r chosen because the system apparently undergoes a weak first-order 
phase transition for this amount of quenched randomness. According to the
finite size scaling arguments conjectured on lattice, for a first-order
phase transition, the valley between the double peaks in the energy histogram
should become deeper and scale as the volume with the increasing lattice size. On the
other hand, the opposite behavior appears in a second-order transition and the
double peak structure in the energy distribution slowly changes to a single
Gaussian as the lattice volume becomes larger.    
From the distributions shown in Fig. 3, one will observe a first-order phase 
transition on a  $64 \times 64$ lattice while a second-order on 
$128 \times 128$ lattice. One will then need to introduce less amount of randomness
~(r smaller than 0.53) to regain the double peak structure in the energy distribution
on the $128 \times 128$ lattice. Apparently, the observed threshold value 
for the amount of 
quenched impurities for the characteristic behavior of the system to change from 
having a first-order phase transition
to  a second-order one is specific to the finite size of the system under
consideration. This threshold gets smaller and smaller as one moves towards
the infinite systems which is what we expect according to the renormalization
group arguments put forward by Hui and Berker  \cite{Hui:1989}.

$\;$ 

We thank  W. Janke and D. Stauffer for fruitful discussions and their valuable 
comments. The hospitality at ZIF, University of Bielefeld where part of this 
work was done is gratefully acknowledged.
This project is partially supported by  Hacettepe University 
Research Fund under the project 95.010.10.003.

$\;$

\pagebreak

\section*{Figure captions}

\begin{description}
\item {Figure 1. Energy histograms for several values of r on 64x64 lattice.}    
                  
\item {Figure 2. Averaged cluster size distributions for several values
of r on 64x64 lattice.}  

\item {Figure 3. Energy histograms for $r=0.53$ on several lattices.}

\end{description}

\pagebreak

\begin{figure}
\psfig{figure=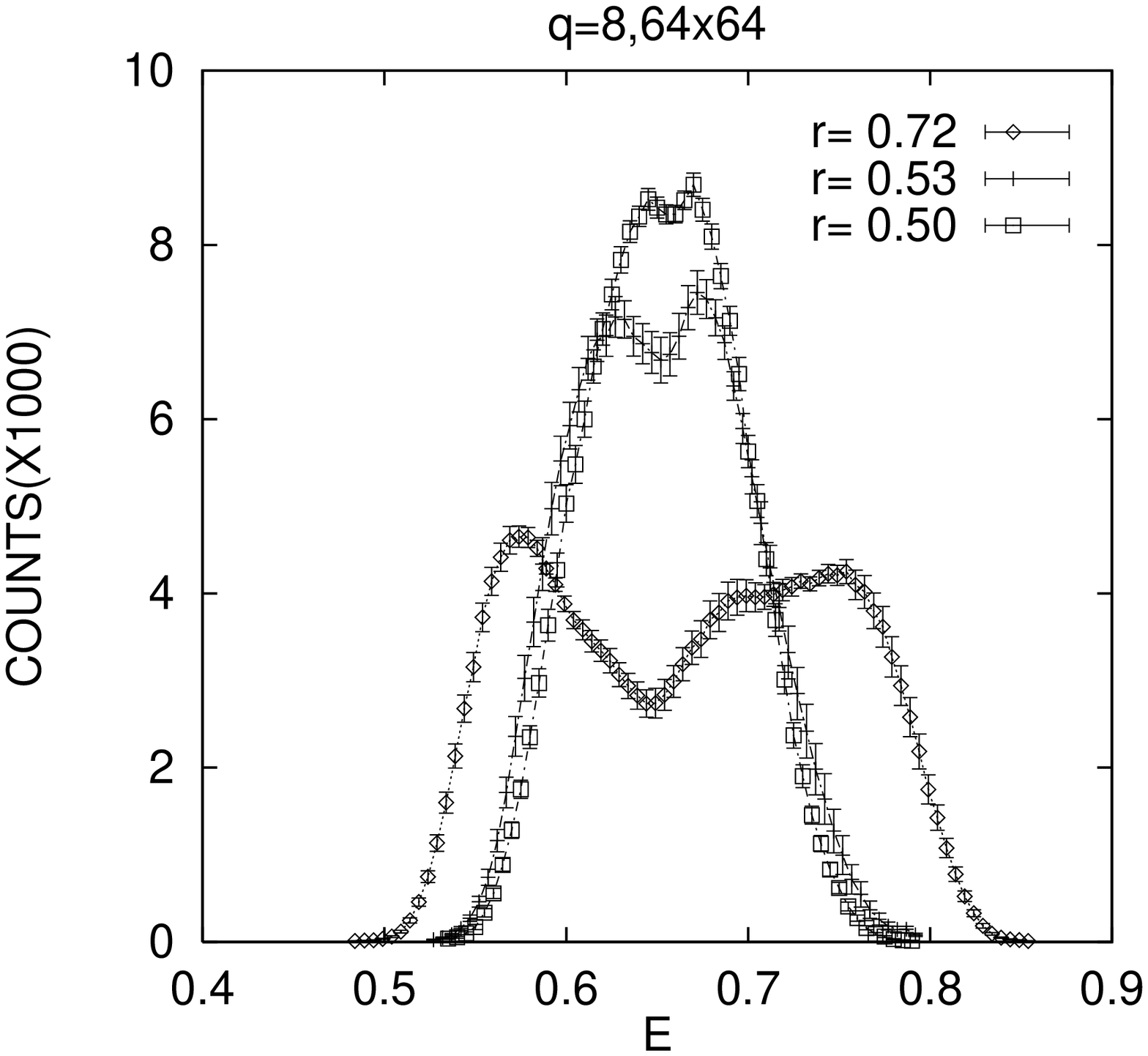,height=8cm,width=12cm,angle=-90}
\caption{}
\end{figure}
\begin{figure}
\psfig{figure=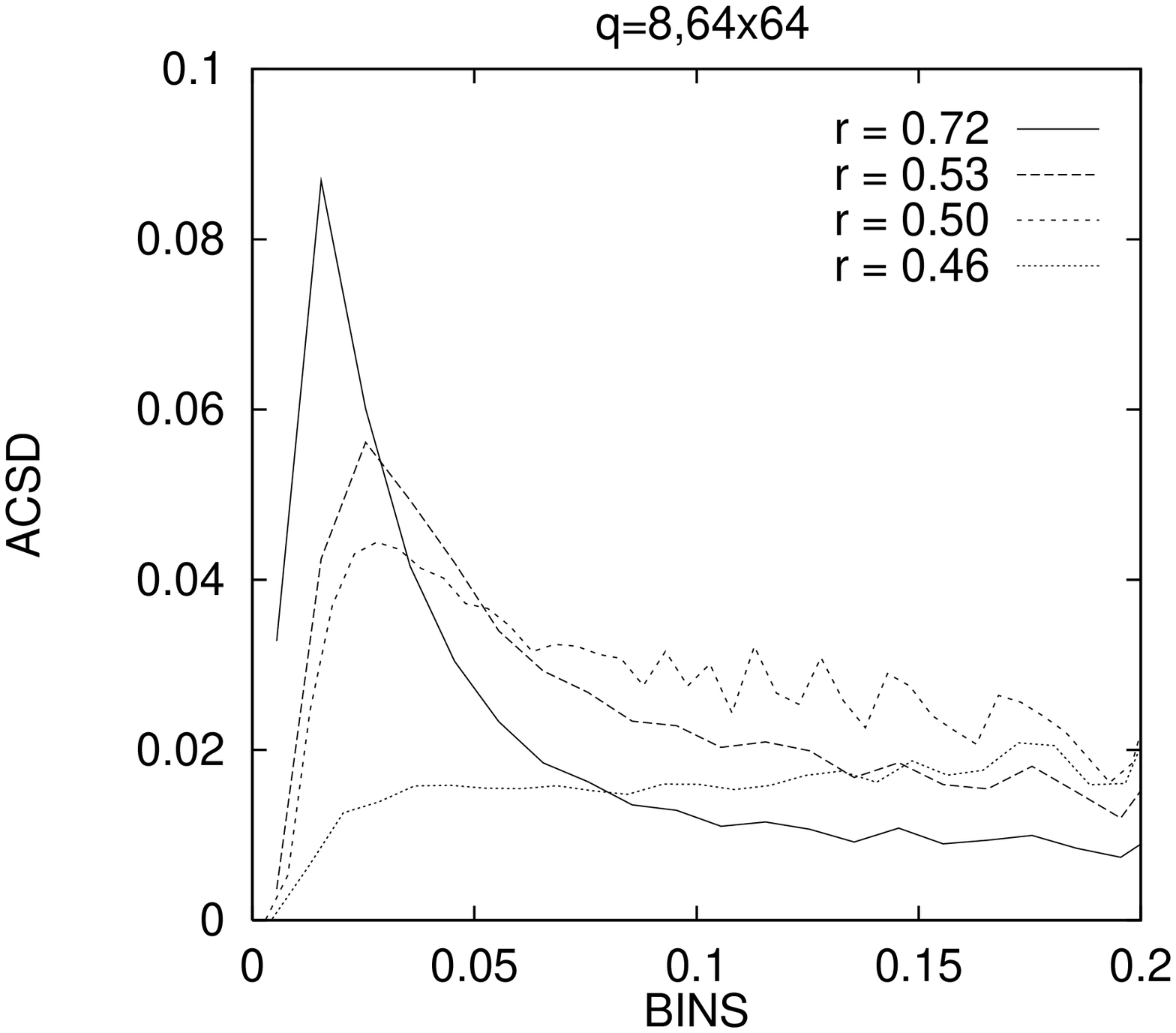,height=8cm,width=12cm,angle=-90}
\caption{}
\end{figure}
\begin{figure}
\psfig{figure=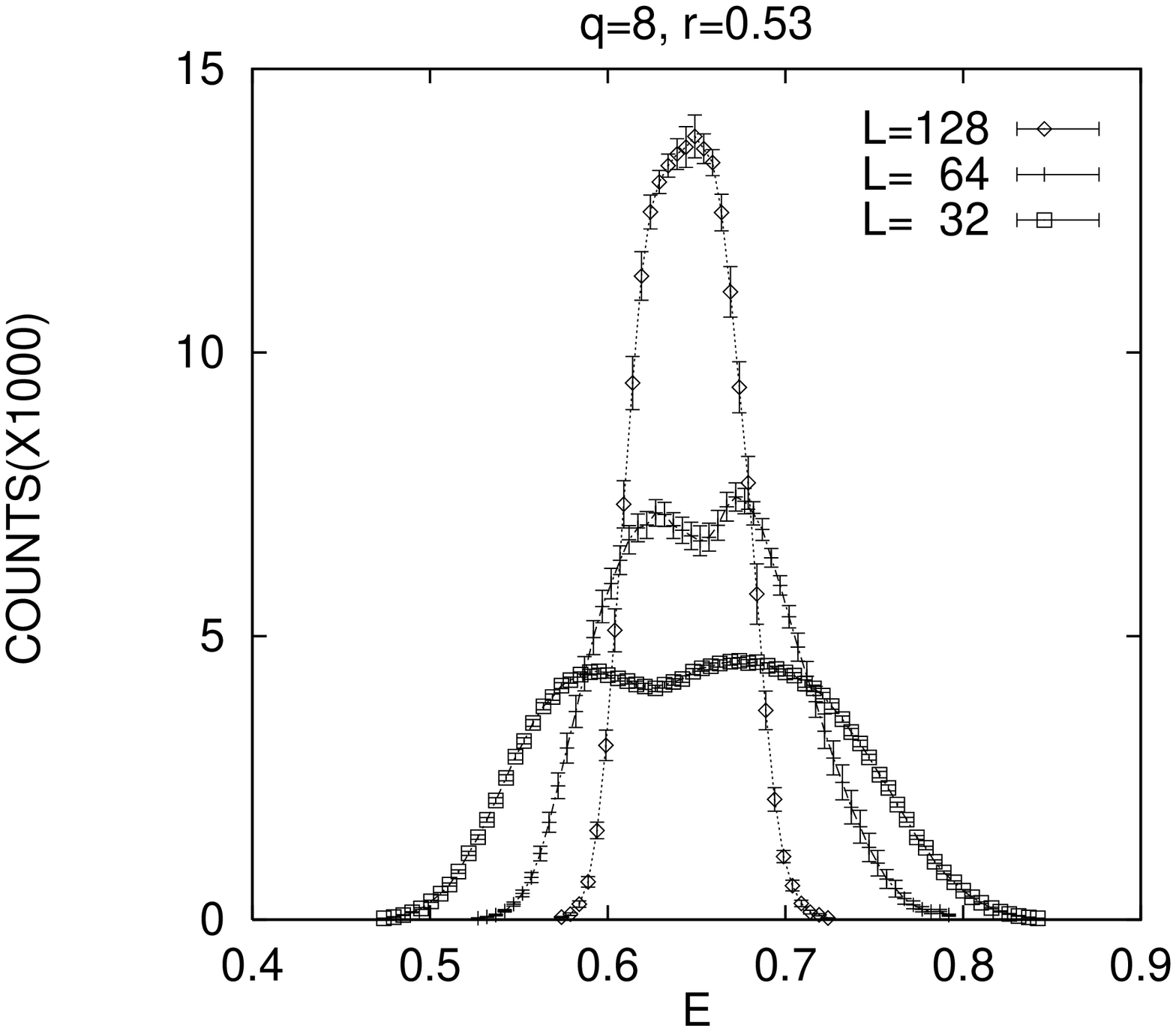,height=8cm,width=12cm,angle=-90}
\caption{}
\end{figure}
\end{document}